\documentstyle[sprocl]{article}
\input psfig
\def\Journal#1#2#3#4{{#1} {\bf #2}, #3 (#4)}

\def\PRL{\em Phys.\ Rev.\ Lett.}

\def\be{\begin{equation}}
\def\ee{\end{equation}}
\def\bea{\begin{eqnarray}}
\def\eea{\end{eqnarray}}

\begin{document}
\title{DATA COMPRESSION FOR CMB EXPERIMENTS}
\author{A.H. JAFFE}
\address{CfPA, 
  301 LeConte Hall, University of California, Berkeley, CA 94720}
\author{L. KNOX \& J.R. BOND} \address{CITA, 60 St. George St., Toronto,
  ON, M5S 3H8 CANADA} 
\maketitle

\abstracts{We discuss data compression for CMB experiments. Although
  ``radical compression'' to $C_\ell$ bands, via quadratic estimators or
  local bandpowers, potentially offers a great savings in computation
  time, they do a considerably worse job at recovering the full
  likelihood than the the signal-to-noise eigenmode method of
  compression.}

We model a CMB observation at a pixel $p=1\ldots N_p$, as
$\Delta_p=s_p+n_p$, where $s$ and $n$ represent the contribution of the
CMB signal and the noise, respectively, to the observation. The signal
is given by $s_p={\cal F}_{p\ell m}a_{\ell m}$; $a_{\ell m}$ is the
spherical harmonic decomposition of the temperature and
${\cal F}$ encodes the beam and any chopping strategy of the experiment.

We also assume that both the signal and noise contributions are
described by independent, zero-mean, gaussian probability distributions,
with correlation matrices given by $\langle s_ps_{p'}\rangle=C_{Tpp'}$
and $\langle n_p n_{p'}\rangle=C_{npp'}$, so $ \langle \Delta_p
\Delta_{p'} \rangle = C_{Tpp'}+C_{npp'}; $ here, $C_T=C_T(\theta)$ is
calculated as a function of $\theta$, the parameters of the theory being
tested in the likelihood function, and the noise matrix can include the
effect of constraints due to, e.g., average or gradient removal. For
Gaussian theories of adiabatic fluctuations, $\theta$ is typically the
cosmological parameters; alternately they could be some set of
phenomological parameters such as the value of the temperature power
spectrum $C_\ell$ in some bands.

With this notation, the likelihood function is
\begin{equation}
  {\cal L}_\Delta(\theta)=P(\Delta|\theta)={\exp\left[-{1\over2}
    \Delta_p\left(C_T(\theta)+C_n\right)^{-1}_{pp'}\Delta_{p'}\right]
    \over(2\pi)^{N_p/2}
    |C_T(\theta)+C_n|^{1/2}}
\end{equation}
Calculating this requires extensive manipulations on the total 
correlation matrix $C_T(\theta)+C_n$ over extensive portions of the 
parameter space. Herein lies the problem: in order to calculate the 
determinant factor in the denominator requires time of $O(N_p^3)$. 

We can state the problem as follows: find some functions of the data,
$f_i\left(\left\{\Delta_p\right\}\right)$ such that the new likelihood,
$ {\cal L}_f(\theta)=P(f|\theta) \simeq{\cal
  L}_\Delta(\theta)=P(\Delta|\theta)$ and ${\cal L}_f$ is ``easier to
calculate'' in some appropriate sense.  The definition of ``$\simeq$''
in this expression is crucial.  If we take it to mean ``having the same
variance,'' the requirement reduces to the definition of ``lossless''
involving the Fisher matrix.\cite{tegmark_optimal} Note
that this requirement will only be adequate very near the maximum of the
distribution, i.e., where the Gaussian approximation is appropriate.
For large data-sets, with high signal-to-noise, this will presumably be
an adequate description; elsewhere (including those parts in the
``tails'' of an otherwise high-S/N experiment's window function that may
be most interesting), it will not necessarily obtain that the derived
confidence limits and parameter estimates will be the same.

If we can find a basis in which the matrices $C_T$ and $C_n$ are
diagonal, the likelihood computation simplifies from matrix
manipulations to $O(N)$ sums and products.  First, we whiten the noise
matrix using the transformation provided by its
``Hermitian square root,'' and apply the same transformation to $C_T$,
diagonalizing this in turn with the appropriate matrix of eigenvectors,
$C_T\to{\rm diag}({\cal E}_k)$, in units of $(S/N)^2$. We then
transform the data into the same basis, ${\Delta}\to{\xi}$, in units of
$(S/N)$.  In this basis, the noise and signal have diagonal
correlations and $\langle\xi_k^2\rangle=1+{\cal E}_k$.  Now, the
likelihood function is a simple product of one-dimensional uncorrelated
gaussians in $\xi_k$.\cite{snmethods}

Modes with low ${\cal E}_{k}$ are linear combinations of the data which
probe the theory poorly. Thus, they are ideal candidates for removal in
a data compression scheme. Removal of these low-S/N modes is the optimal
way to compress the data: for a given number of modes, it removes the
most noise and least signal.  Of course, the modes will change as the
shape of the theory used for the covariance matrix $C_T$ changes. In
that case, we choose some fiducial theory (here, standard, untilted CDM)
that represents the data moderately well, calculate its S/N-eigenmodes,
and choose some number of modes to retain after compression. For other
theories, these modes will not be optimal---we will have removed more
signal and less noise. Nonetheless, this method does quite well even far
from the fiducial theory, as we see in the Figure, which shows the
likelihood for a parameterization of a standard CDM power spectrum with
amplitude, $\sigma_8$, and a scalar tilt, $n_s$ (so the primordial
$P(k)\propto k^{n_s}$).  Unfortunately, finding this basis is $O(N^3)$;
even performing this operation once for a megapixel dataset will be
prohibitively expensive.

\begin{figure}[htbp]
  \centerline{\psfig{file=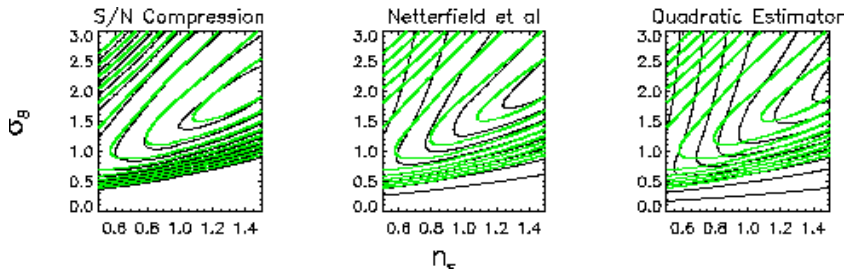,height=1.5in,width=4.5in}}
  \caption{Contours of constant likelihood ratio (at $\delta2\ln{\cal 
      L}=1,4,9,\ldots$) for the SK94-5 data, parameterized by $\sigma_8$
    and $n_s$, the primordial spectral slope, for different
    methods of data compression. The light-colored contours
    in each panel show the uncompressed likelihood.}
  \label{fig:SK95Compress}
\end{figure}

Most experiments report their results in the form $[{\hat{\cal
    C}}_\ell\pm\delta {\cal C}_\ell]$; this encourages the use of a
simple curve-fitting approach to parameter estimation\cite{lineweaver}:
form the obvious quantity 
$\chi^2(\theta)=\sum_\ell{\left[{\hat{\cal C}}_\ell-{\cal
      C}_\ell(\theta)\right]^2 /\left(\delta{\cal C}_\ell\right)^2}$
where ${\cal C}_\ell(\theta)$ is the predicted power spectrum for the
theoretical parameters $\theta$. In practice, the power spectrum is
usually reported as a flat bandpower over some $\ell$ band with an
appropriate window function, but the procedure remains the same.

Now, just do the usual fast $\chi^2$-minimization for the parameters.
This is a very ``radical'' approximation to the full likelihood: it
assigns an uncorrelated gaussian distribution to the power spectrum:
${\cal L}(\theta)\propto\exp(-\chi^2/2)$. 
In the figure, we show confidence intervals for the SK95 experiment with
the full likelihood and compare them to those obtained with 1) the S/N
compression discussed above; 2) $\chi^2$ using flat bandpowers as
reported by Netterfield et al;\cite{nett}; 3) $\chi^2$ using a
quadratic estimator of $C_\ell$ in bands of $\ell$, modified somewhat to
minimize covariance\cite{quad_Cl}. The peaks are nearby, and the
contours are similar in the ``amplitude'' direction ($\sigma_8$), but
less good in the ``shape'' ($n_s$) direction.  Far from the peak, all of
the $\chi^2$ methods do quite poorly. The location of the peak is
determined by the gross shape of the power spectrum, so the details of
the calculation do not matter; this is especially true when combining
the results of experiments probing very different scales, such as
COBE/DMR and SK95. Away from the peak, the shape of the spectrum within
the experimental windows, the covariances of the errors, and the
non-Gaussian shape all contribute to the determination of which theories
are more highly disfavored.




\section*{References}
\begin{thebibliography}{9} 
\bibitem{tegmark_optimal} M.\ Tegmark, {\tt astro-ph/9611174}, 
  {\sl Ap.\ J.}, submitted, {1996}.
\bibitem{snmethods} J.R.\ Bond \& A.\ Jaffe,
  {\sl\PRL}, {submitted}, {1997}; A.\ Jaffe \& J.R.\ Bond, in
  preparation, 1997.
\bibitem{lineweaver} C.\ Lineweaver, these proceedings.\  
\bibitem{nett} C.B.\ Netterfield, M.J.\ Devlin, N.\ Jarosik, L.\ Page, \&
  E.J.\ Wollack,  \Journal{\sl Ap.\ J.}{474}{47}{1997}.
\bibitem{quad_Cl} L.\ Knox, J.R.\ Bond \& A.\ Jaffe, these proceedings;
  M.\ Tegmark \& A.\ Hamilton,  these proceedings.
\end{thebibliography}
\end{document}